\documentstyle[sprocl,epsfig]{article}

\def\be{\begin{equation}}
\def\ee{\end{equation}}
\def\bea{\begin{eqnarray}}
\def\eea{\end{eqnarray}}

\def\lsim{\raise0.3ex\hbox{$<$\kern-0.75em\raise-1.1ex\hbox{$\sim$}}}
\def\gsim{\raise0.3ex\hbox{$>$\kern-0.75em\raise-1.1ex\hbox{$\sim$}}}

\begin{document}

\title{DECONFINEMENT AND CHIRAL SYMMETRY RESTORATION}

\author{Frithjof Karsch}
\address{Fakult\"at f\"ur Physik\\
           Universit\"at Bielefeld, D-33615 Bielefeld, Germany}

\maketitle\abstracts{ 
We discuss the critical behaviour of strongly interacting matter
close to the QCD phase transition. Emphasis is put on a presentation 
of results from lattice calculations that illustrate deconfining
as well as chiral symmetry restoring features of the phase transition.  
We show that both transitions coincide in QCD while they fall apart in
an SU(3) gauge theory coupled to adjoint fermions. We also discuss
some results on deconfinement in quenched QCD at non-zero baryon
number.}

\section{Introduction}

The interest in analyzing the properties of QCD at non-zero temperature 
is twofold. On the one hand it is the goal to reach a quantitative description
of the behaviour of matter at high temperature and density. This does
provide important input for a quantitative description of experimental 
signatures for the occurrence of a phase transition in heavy ion 
collisions and should also help to understand better the phase transitions that
occurred during the early times of the evolution of the universe.
Eventually it also may allow to answer the question whether a quark-gluon
plasma can exist in the interior of dense neutron stars.  
For this reason one would like
to reach a quantitative understanding of the QCD equation of state, 
determine critical parameters such as the critical temperature and 
the critical energy density and predict the modification of basic
hadron properties (masses, decay widths) with temperature and density.
On the other hand the analysis of a complicated quantum field theory
at non-zero temperature can also help to improve our understanding
of its non-perturbative properties at zero temperature. The introduction
of an external control parameter (temperature) allows to observe the response
of different observables to this and may provide a better understanding
of their interdependence \cite{Wilczek}. In particular,
one would like to clarify the role of confinement and chiral symmetry
breaking for the QCD phase transition. 

In which respect is the QCD
phase transition deconfining and/or chiral symmetry restoring? 
In the next section we will address this question 
and will present some basic results on the QCD equation of state
and critical parameters at the transition point obtained from lattice
QCD. In section 3 we will discuss deconfinement and chiral symmetry
restoration in an SU(3) gauge theory with adjoint fermions.   
First exploratory results on deconfinement in QCD at
non-zero baryon number will be discussed in section 4 and in section 5 we 
give our conclusions. 
Throughout this write-up we will as far as possible try to avoid going into 
details of the actual lattice calculations. Basics concepts of the 
lattice formulation of QCD relevant for finite temperature calculations
and in particular for the discussion of deconfinement and chiral
symmetry restoration can be found for instance in
Refs.~2 and 3.

\section{The QCD phase transition}

Two properties of QCD explain the basic features of the observed
spectrum of hadrons -- confinement and chiral symmetry breaking.
While the former explains why we observe only colourless states in the
spectrum the latter describes the presence of light Goldstone particles,
the pions. The confining property of QCD manifests itself in the large 
distance behaviour of the heavy quark 
potential. At zero temperature the potential rises linearly at large
distances, $V_{\bar{q}q} (r) \sim \sigma r$, where $\sigma \simeq 
(420~{\rm MeV})^2$ denotes the string tension, and forces the quarks and 
gluons to be confined in a {\it hadronic bag}. Chiral symmetry breaking 
leads to a non-vanishing quark anti-quark condensate, 
$\langle \bar{q}q  \rangle \simeq (250~{\rm MeV})^3$, in the vacuum.
Inside the hadron bag, however, the condensate vanishes. 
At high temperatures the individual hadronic bags are expected to merge to a 
single large bag, in which quarks and gluons can move freely. This bag
picture provides some intuition for the occurrence of 
deconfinement and chiral symmetry restoration.
A priory it is, however, not evident that confinement and the broken chiral
symmetry have to get lost at the same temperature.
It has been speculated that two distinct phase transitions leading
to deconfinement at $T_d$ and chiral symmetry restoration at $T_\chi$
could occur in QCD \cite{Shuryak}. General arguments about the scales
involved suggest that $T_d \le T_\chi$.
 
\begin{figure}
\begin{center}
\epsfig{
         file=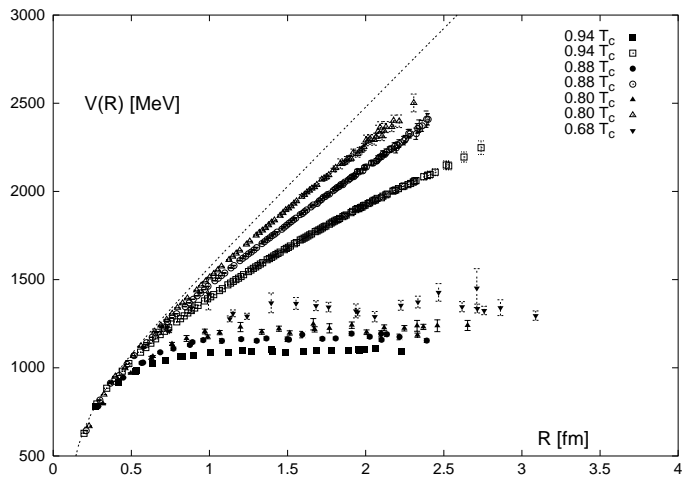,height=65mm}
\end{center}
\caption{The heavy quark potential in 2-flavour QCD with quarks
of mass $m/T=0.15$ extracted from calculations on a $16^3\times 4$ 
lattice and in quenched QCD on $32^3\times 4$ lattices. The dashed line
shows results for the potential at $T\simeq 0$.
\label{fig:potential}
}
\end{figure}

In fact, the discussion of confinement and deconfinement in terms of the
heavy quark potential as presented above makes sense only in the 
limit of heavy quarks. For light quarks the spontaneous creation of 
quark anti-quark pairs from the vacuum leads to
a breaking of the ``string'' between static quark sources, {\it i.e.} 
the potential tends to a constant
value for $r \rightarrow \infty$. In particular at large temperature
the distinction between confinement and deconfinement thus seems to  
become a qualitative one. This is evident from a comparison of heavy
quark potentials calculated in QCD with light quarks as well as in the quenched
limit at temperatures close but below the phase transition \cite{DeTar}.
The potential shown in Figure~\ref{fig:potential} has been obtained  
from a calculation of 
expectation values of the Polyakov loop correlation function
\begin{equation}
\exp{\biggl( -{V_{\bar{q}q} (r,T) \over T}\biggr) } = \langle 
{\rm Tr} L_{\vec{x}} {\rm Tr} L^{\dagger}_{\vec{y}} \rangle \quad,\quad 
r=|\vec{x}-\vec{y}|
\label{poly}
\end{equation}  
where $ L_{\vec{x}}$ and $L^{\dagger}_{\vec{y}}$  represent a static quark 
and anti-quark pair located at the spatial 
points $\vec{x}$ and $\vec{y}$, respectively\footnote{Further details on the
definition of this observable is given for instance in Ref. 2
}. 

At large distances the Polyakov loop correlation function, Eq.~\ref{poly},
approaches $|\langle L \rangle|^2$, where $L$ is given by
$ L  = N_{\sigma}^{-3} \sum_{\vec{x}} {\rm Tr}L_{\vec{x}}$. The Polyakov
loop expectation value $\langle L \rangle$ thus 
reflects the behaviour of the heavy quark potential at large distances. 
A sudden rise in $\langle L \rangle$ indicates that the heavy 
quark potential flattens already at even shorter distances; the magnitude 
of $\langle L \rangle$ thus still signals the 
transition from a confining to a deconfining thermal medium. This transition
also manifests itself in the
occurrence of a peak in the Polyakov loop susceptibility,
\begin{equation}
\chi_L = \langle L^2  \rangle -  \langle L  \rangle^2~~ \quad . 
\end{equation}
The temperature dependence of chiral properties of QCD on the other
hand becomes visible directly through the temperature dependence
of the chiral condensate, $\langle \bar{\psi}\psi  \rangle$ as well as its
derivative with respect to the quark mass, the chiral susceptibility,   
$\chi_{m}$,
\begin{equation}
\chi_{m} = {\partial \over \partial m_q} \langle \bar{\psi}\psi 
\rangle
\quad .
\label{sus}
\end{equation}

\begin{figure}
\vspace{-2.0cm}
\begin{center}
\hspace{-1.2cm}
\epsfig{file=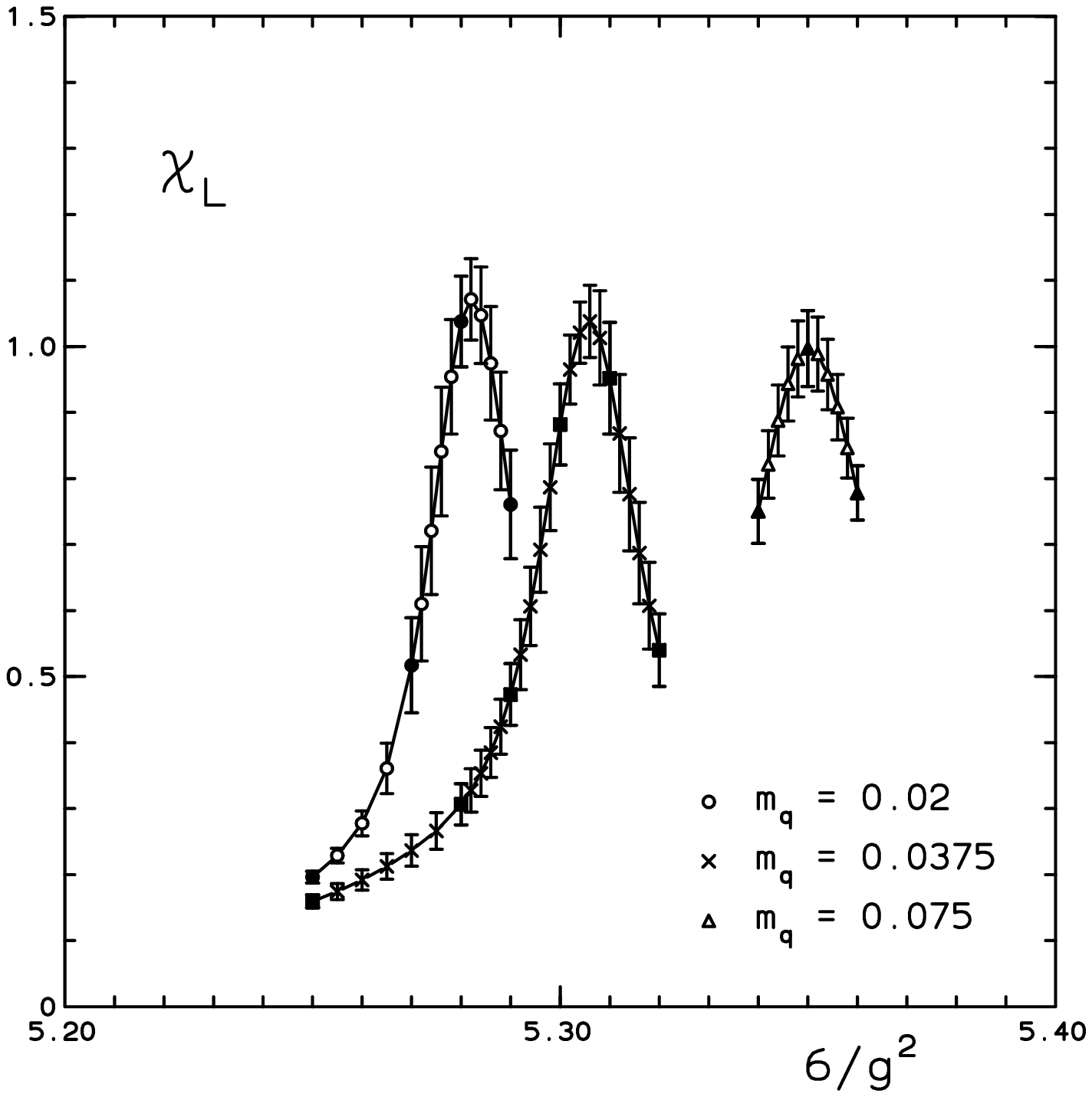,width=73mm} \hspace{-1.8cm}
\epsfig{file=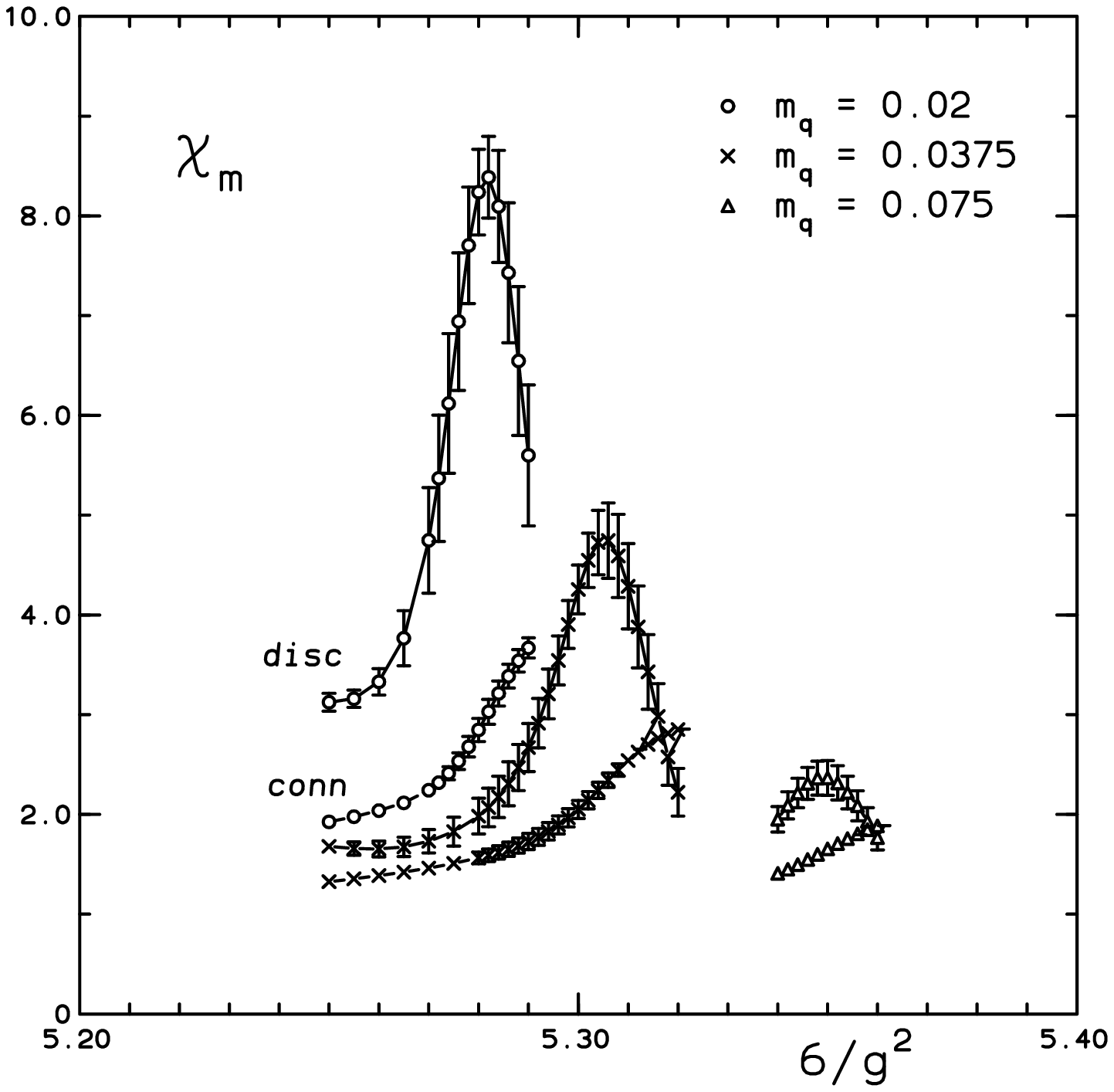,width=73mm}
\end{center}
\vspace{-2.0cm}
\caption{Polyakov loop and chiral susceptibilities versus $\beta=6/g^2$
in 2-flavour QCD for several values of the quark mass.
\label{fig:susmass}
} 
\end{figure}

Results from a calculation of these observables for QCD with two light
quarks \cite{KarL94} are shown in Figure~\ref{fig:susmass} as a function of 
the lattice bare coupling $\beta=6/g^2$. Evidently the peak in 
both susceptibilities is located at the same value of the
(pseudo)-critical couplings, $\beta_{c}(m_q)$. This indicates that
the transition to a deconfined, chirally symmetric phase occurs at the 
same temperature. These results provide the basic evidence for the 
existence of a single phase transition in QCD. 

In fact, the finite peak heights of the susceptibilities shown in  
Figure~\ref{fig:susmass} as well as
the rapid but smooth variation of $\langle L  \rangle$ and
$\langle \bar{\psi}\psi \rangle$ itself do not
yet correspond to a {\it true} phase transition. This would be
signaled by diverging susceptibilities and is expected to occur
only in the infinite volume and zero quark mass limit\footnote{This is
correct only when
the phase transition is second order. In the case of a first
order transition the susceptibilities would diverge already for masses below 
a certain 
non-zero quark mass.}.  In this respect it is interesting to note 
that the chiral susceptibility grows rapidly with decreasing
quark mass while the Polyakov loop susceptibility shows little variation
with $m_q$ in the quark mass regime covered in Figure~\ref{fig:susmass}.
In fact, when compared to the heavy quark mass regime the peak in the Polyakov
loop susceptibility has dropped substantially \cite{Peikertp}. 

\begin{figure}[htb]
\begin{center}
\epsfig{file=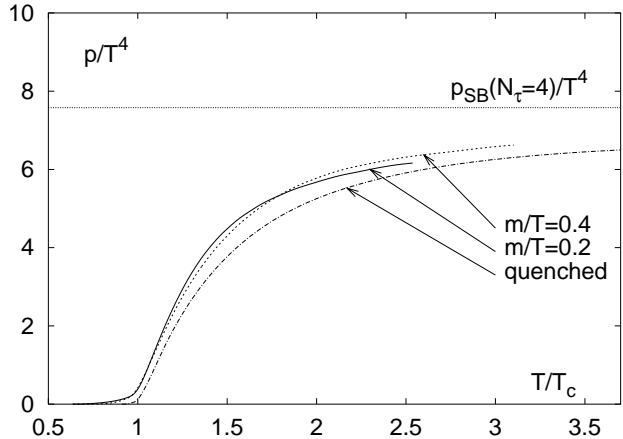,width=90mm}
\end{center}
\caption{The pressure in units of $T^4$ versus $T/T_c$ for a pure SU(3)
gauge theory (quenched) and four flavour QCD with two different values
of the quark mass ($m/T=0.2,~0.4$). The numerical results for four flavour QCD
have been rescaled by the corresponding ratio of ideal gas values, 
$p_{\rm SB} (n_f=4)/p_{\rm SB} (n_f=0)$.
\label{figpressure}
}
\end{figure}

We conclude that observables directly related to chiral symmetry
restoration show critical behaviour in the zero quark mass limit, while
observables related to deconfinement do not seem to become singular
in this limit. In this respect it may be justified to call the QCD phase 
transition a {\it chiral phase transition}. 
However, this is only one feature of the transition. The 
deconfining aspect of the transition is particularly evident when 
looking at the behaviour of bulk thermodynamic observables like the
energy density or the pressure. Asymptotically, for $T\rightarrow \infty$,
these quantities approach the value of an ideal gas and thus directly
reflect the number of light degrees of freedom in the system. For
QCD with $n_f$ light (massless) quarks and anti-quarks as well as
$N_c^2-1$ gluons ($N_c=3$) the pressure and energy density approach the 
Stefan-Boltzmann value,
\begin{equation}
{p_{\rm SB} \over T^4} = {1\over 3} {\epsilon_{\rm SB} \over T^4}
={\pi^2 \over 45}\biggl( N_c^2-1 +{7\over 4} N_c n_f \biggr)~~~,
\label{SB}
\end{equation}
which directly counts the relevant number of light degrees of freedom.
Above $T_c$ these degrees of freedom get liberated which is reflected
in a rapid rise of the energy density and the pressure. In 
Figure~\ref{figpressure} we show a comparison of the temperature dependence
of the pressure calculated in the SU(3) gauge theory \cite{Boydxx} 
($n_f=0$) and four flavour QCD \cite{Joswig} ($n_f=4$). In both cases
the high temperature limits differ by more than a factor of three. 
Nonetheless, when normalized 
to the corresponding Stefan-Boltzmann values the pressure $p/p_{\rm SB}$
shows a similar temperature dependence. Bulk 
thermodynamic observables like the pressure thus show that the light
partonic degrees of freedom indeed get liberated at $T_c$. Their sudden 
increase reflects the onset of deconfinement.

\section{SU(3) gauge theory with adjoint fermions}

To some extent the existence of a single phase transition in QCD may not be 
too surprising. After all the QCD Lagrangian has only one global symmetry --
chiral symmetry, which at vanishing
quark mass is spontaneously broken. Only in the pure gauge limit,
{\it i.e.} the limit of infinitely heavy quarks there exists another exact
symmetry, the Z(3) center symmetry, which is related to confinement 
in the literal sense that $V_{\bar{q}q}(r,T)$ approaches infinity for
$r \rightarrow \infty$. The Z(3) symmetry
does get spontaneously broken at the critical temperature of the SU(3) gauge 
theory. For all finite values of the quark mass it is, however, explicitly 
broken. The importance of the realization of an exact center symmetry in the
Lagrangian for the
existence of a {\it true} phase transition related to deconfinement 
can be analyzed in a QCD-related model like the SU(3) gauge theory
with fermions in the adjoint rather than in the fundamental 
representation (aQCD) \cite{kogut,Luetgemeier}.
The lattice formulation of aQCD is obtained from that of ordinary QCD 
by replacing the three-dimensional representation
of the gauge link matrices, $U^{(3)}$, in the 
fermionic part of the action by the corresponding eight-dimensional
representation, $U^{(8)}$,
\begin{equation}
S_{{\rm aQCD}} = S_G + S_F(U^{(8)})\quad {\rm with} \quad U^{(8)}_{a,b} =
{1\over 2} {\rm Tr} U^{(3)}\lambda_a U^{(3)}\lambda_b~~.
\end{equation}

\begin{figure}
\vspace{-1.0cm}
\begin{center}
\hspace{-0.5cm}
\epsfig{file=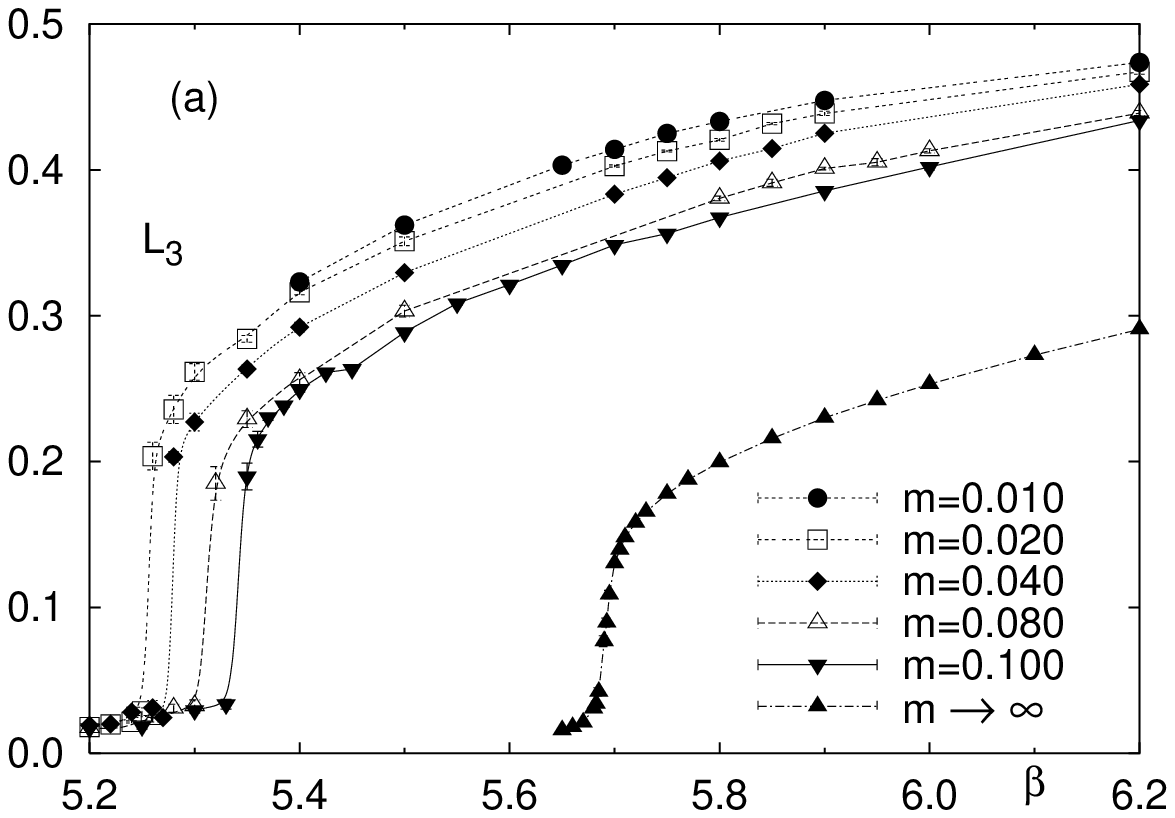, width=64mm}
\hspace{-0.7cm}
\epsfig{file=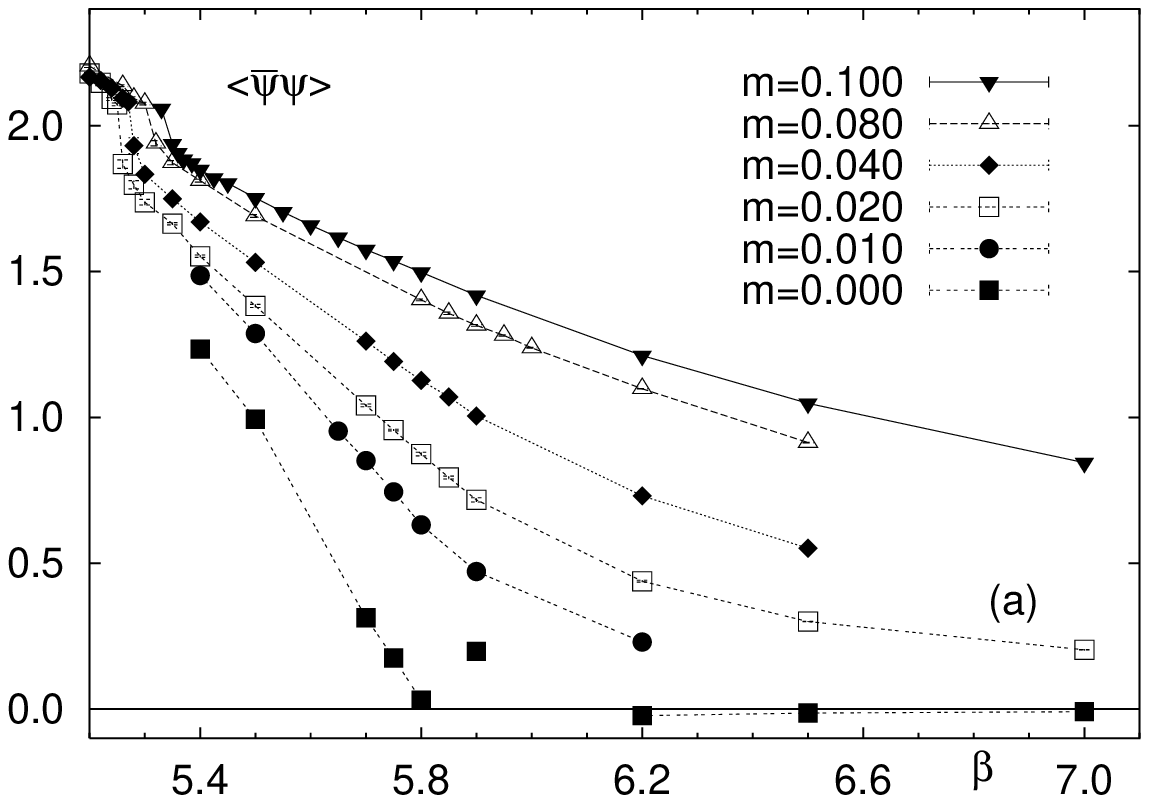, width=64mm}
\end{center}
\caption{Polyakov loop expectation value, $L_3$, in the fundamental  
representation obtained from simulations on $8^3\times 4$
lattices with adjoint fermions of various masses (left) and the fermion 
condensate (right) obtained with the same simulation parameters.
\label{fig:adjoint}
} 
\end{figure}

At zero temperature this theory has a broken chiral symmetry and an exact 
Z(3) center symmetry, {\it i.e.} for all values of the adjoint fermion mass 
it is strictly confining for fundamental charges which are used to probe the
heavy quark potential. What is less obvious though is that the Z(3) 
symmetry plays any role for the deconfinement of the dynamical degrees of
freedom participating in the thermodynamics, in particular for the adjoint 
quarks which are blind to the center symmetry\footnote{The potential between 
static charges in the adjoint representation is not
confining; at large distances string breaking occurs.}. 
In Figure~\ref{fig:adjoint}
we show the Polyakov loop expectation values for fundamental and
adjoint charges as well as the fermion condensate. It is evident that
there are two distinct phase transitions, {\it i.e.} deconfinement occurs 
before chiral symmetry restoration ($T_\chi \simeq 8 T_d$) \cite{Luetgemeier}.

\begin{figure}
\vspace{-1.0cm}
\begin{center}
\hspace{-1.0cm}
\epsfig{file=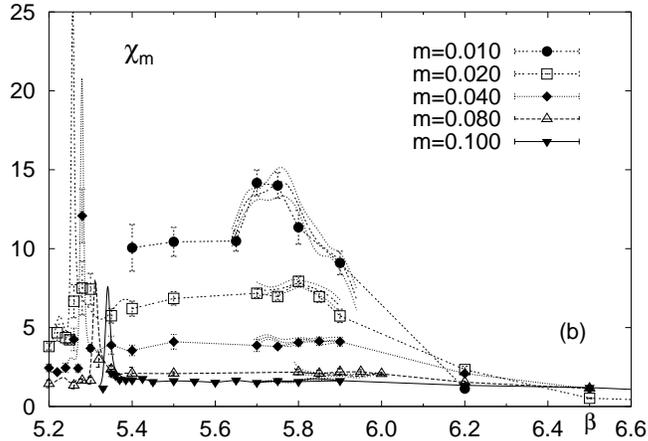, width=90mm}
\end{center}
\caption{Polyakov loop susceptibility of aQCD versus $\beta$ for several
values of the fermion mass.
\label{fig:aqcdsus}
} 
\end{figure}

Contrary to QCD the SU(3) gauge theory with adjoint quarks has an
intermediate phase where confinement is lost but chiral symmetry remains
broken. It is apparent from  Figure~\ref{fig:adjoint} that in this
phase the chiral condensate is much more sensitive to changes of the
fermion mass than in the confined phase. In fact, for $T_d < T <T_\chi$  
the leading fermion mass dependence of the chiral condensate is  
\begin{equation}
\langle \bar{\psi} \psi \rangle = a_0 + a_1 m^{1/2} + {\cal O}(m)~~,
\label{gold}
\end{equation}
whereas below $T_d$ the leading correction only starts at ${\cal O}(m)$. 
This is even more evident from the chiral susceptibility of aQCD
shown in Figure~\ref{fig:aqcdsus}. In the intermediate phase $\chi_m$ shows
a strongly quark mass dependent plateau which diverges like $m^{-1/2}$.
The occurrence of such a divergence in the chiral susceptibility below
but close to to $T_\chi$ has been expected. It was shown to exist in
$3d$ sigma model where it arises from fluctuations of Goldstone modes
in the unbroken directions \cite{Zia}. We thus find that universality
indeed links the chiral properties of (3+1) dimensional aQCD below $T_\chi$
and $3d$ sigma model. To establish this relation also directly at $T_\chi$
seems to be more difficult. Here a more rapid divergence, 
$\chi_m \sim m^{1/\delta-1} \sim m^{-0.8}$, is expected. Although the 
simulation results shown in Figure~\ref{fig:aqcdsus} indicate such a more
rapid divergence at $T_\chi$, the corresponding peak in $\chi_m$ shows up
only for rather small quark masses and a reliable determination of critical 
indices is not yet possible.

A similar behaviour is expected to occur in QCD. So far the studies of
chiral properties at the QCD phase transition, however, did not yield the
critical indices expected on the basis of universality 
arguments \cite{KarL94,JLQCD}.
From the study of chiral symmetry breaking in aQCD one might conclude 
that the occurrence of additional subleading singularities as well as the
strong influence of confinement on the quark mass dependence of chiral
observables do make a more complex analysis of chiral observables in QCD 
necessary that also takes into account subleading dependences on the quark
mass.
 
\section{Deconfinement in quenched QCD at non-zero baryon number}

The formulation of QCD at non-zero baryon number density is known to
lead to substantial numerical difficulties as the Euclidean
path integral formulation no longer possesses a positive integration 
kernel. As a consequence little is known from lattice calculations with
light quarks about deconfinement and chiral symmetry restoration in 
dense hadronic matter. We want to discuss here a less demanding, although
still complicated problem -- the thermodynamics of gluons in the background
of static quark sources. This is the quenched limit of QCD at non-zero
baryon number. Contrary to the conventional approach we will introduce
a non-zero baryon number rather than a non-zero chemical potential and
perform the heavy quark mass limit of QCD\footnote{The heavy quark mass
limit at non-zero chemical potential is discussed in Ref.~19.}.  
This is an alternative 
approach to finite density QCD \cite{Redlich}. Both formulations are related 
through a Fourier transformation of the QCD partition function with a complex
chemical potential, $\mu=i\phi$,
\begin{equation}
Z(B,T,V) = {1\over 2\pi}
\int_0^{2\pi}  {\rm d}\phi\ {\rm e}^{-iB\phi}\ Z(i\phi,T,V)~~.
\label{density} 
\end{equation} 
In the heavy quark mass (quenched) limit also the finite baryon number
formulation does not lead to a strictly positive integrand in
the partition function. Simulations, however, can be performed with the
absolute value of the integrand and its sign can be included in the 
calculation of expectation values. In this way thermodynamic observables  
can be analyzed in complete analogy to the case of vanishing baryon 
number \cite{Engels99}.

\begin{figure}
\vspace{-1.0cm}
\begin{center}
\hspace{-0.5cm}
\epsfig{file=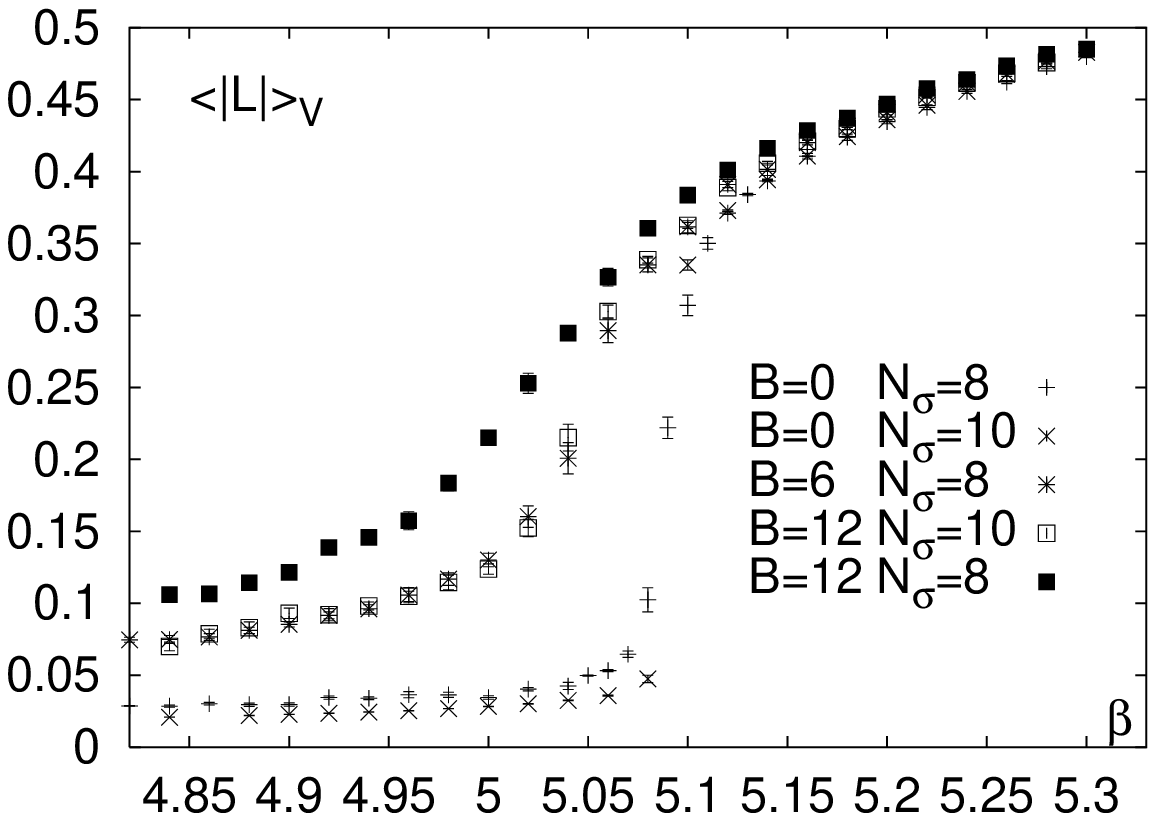,width=63mm}
\hspace{-0.5cm}
\epsfig{file=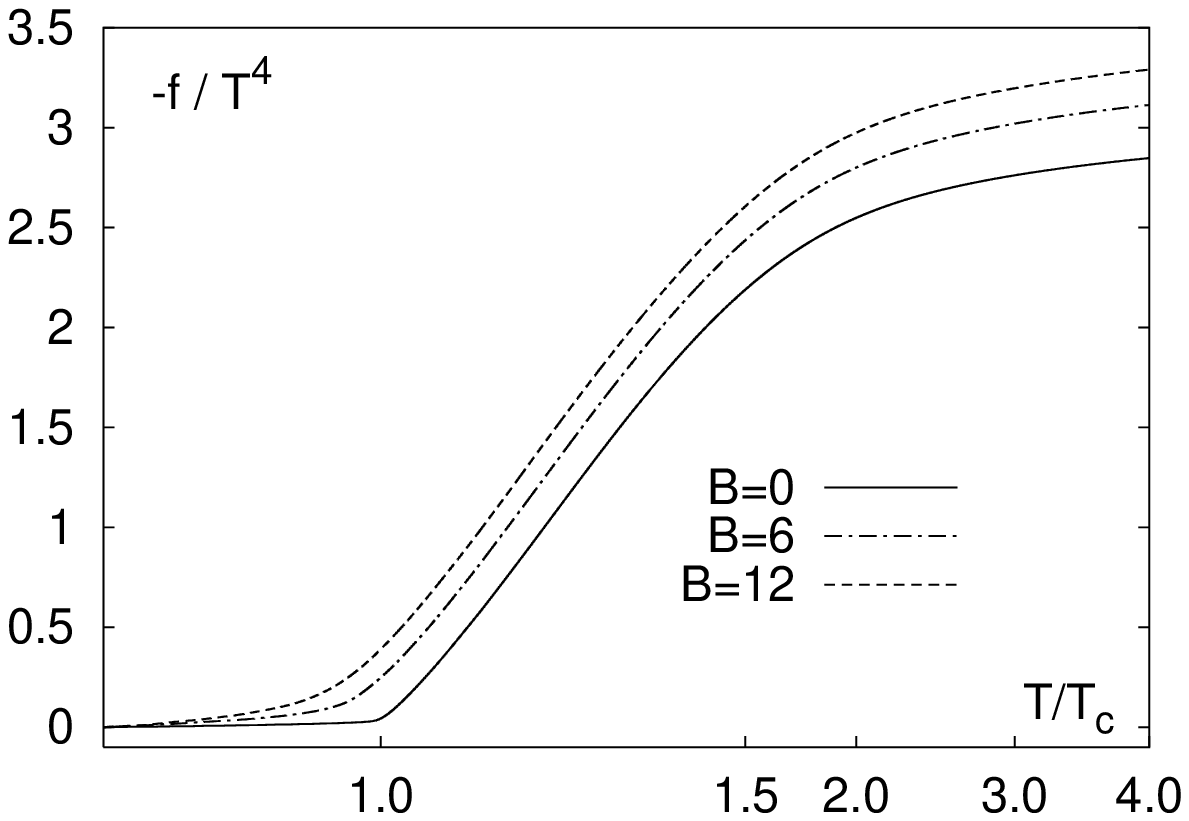,width=63mm}
\end{center}
\caption{Expectation value of $|L|$ (left) and the negative of
the free energy density in units of $T^4$ (right) versus $\beta$ for 
$B=0,~6$ and 12 on lattices of size $8^3 \times 2$ and $10^3 \times 2$. 
\label{fig:baryon}
}
\end{figure}

In the previous sections we have seen that the Polyakov loop expectation
value provides basic information about the occurrence of deconfinement.
Its behaviour is linked to the long distance behaviour of the heavy quark
potential and also reflects the restoration/breaking of the 
global Z(3) symmetry of the QCD action. On the first sight these two
features seem to lead to contradictory expectations for the behaviour
of $\langle L \rangle$ at non-zero baryon number.  
The non-zero baryon number formulation of QCD is known to project
onto the zero triality sector, {\it i.e.} the partition function 
generated through the Fourier transformation in Eq.~\ref{density}
has an integrand that is explicitly Z(3) symmetric. One 
thus might expect that the Polyakov loop expectation value is again
an order parameter for confinement as it is in the $B=0$ case. However, 
for $B>0$ there are already many quarks present in the thermal medium (even
in the quenched limit). Static quark anti-quark sources
used to probe the heavy quark potential can thus recombine with the already present
static quarks and will lead to string breaking even in the low temperature
phase. One thus would expect that the Z(3) symmetry has to be broken for $B>0$ 
and that $\langle L \rangle > 0$ for all
temperatures. This is confirmed by the numerical simulation on 
$8^3 \times 2$ and $10^3\times 2$ lattices shown in Figure~\ref{fig:baryon}. 
We also note that
only for $B=0$ the deconfinement transition leads to a discontinuity
in the order parameter. For $B>0$ we find a transition region that broadens
with increasing $B$. It separates a phase with 
small but non-zero, strongly $B$-dependent  Polyakov loop expectation 
values and a phase with a large approximately $B$-independent expectation 
value. In how far the transition region is a mixed phase related to
a first order deconfinement transition remains to be analyzed in more detail.  

With increasing $B$ the Polyakov loop expectation value is no longer
an unambiguous indicator for deconfinement. It will show little
variation with temperature. As we have discussed previously deconfinement
should, however, clearly be visible in the temperature dependence
of bulk thermodynamic observables. We thus expect to find e.g. a rapid
change in the free energy density that reflects the liberation of many
gluonic degrees of freedom. This is indeed observed in our simulations
as can be seen in Figure~\ref{fig:baryon}.  With increasing baryon number 
density the onset of deconfinement reflected by a rapid rise in $-f/T^4$
is shifted towards smaller temperature. For the $B=12$ data shown in 
Figure~\ref{fig:baryon} the baryon number density in the critical 
temperature interval is about nuclear matter density.
The shift in temperature visible here between the free energies for $B=0$ 
and $B=12$ is about $15\%$.

\section{Conclusions} 
Strongly interacting matter does undergo a single phase transition
at finite temperature, which is deconfining and chiral symmetry restoring.
We have discussed some aspects of the transition which show
its chiral symmetry restoring as well as deconfining properties.

The chiral order parameter and its susceptibility show a strong
quark mass dependence which signals the occurrence of a phase
transition in the zero quark mass limit. However,
the details of this quark mass dependence 
are in the case of two-flavour QCD so far not in agreement with the 
expected universal behaviour of the 3-dimensional $O(4)$ symmetric 
sigma-models. Nonetheless the transition seems to be continuous, at 
least in the quark mass regime so far accessible to numerical 
calculations ($m_q/T \gsim 0.04$) there are no indications for a first 
order transition. 

The analysis of an SU(3) gauge theory with adjoint fermions shows
that chiral observables exhibit universal scaling behaviour close to 
$T\chi$ as expected from the analysis of $3d$ sigma models. 
Below the deconfinement transition temperature, $T_d$, confinement dominates
and these singularities are no longer visible. This indicates 
that even close to $T\chi$ the chiral sector of QCD may be influenced 
strongly by confinement as well as by contributions from subleading chiral 
singularities.

The exploratory analysis of quenched QCD at non-zero baryon number shows
that a deconfining transition occurs also in this case. The onset
of deconfinement is shifted to lower temperatures. Although the 
partition function possesses an exact Z(3) symmetry the Polyakov loop 
is non-zero at all temperatures for non-zero baryon number. This indicates
that the heavy quark potential will show string breaking in the 
low temperature phase. 

\vfill\eject
\bigskip\noindent
{\bf Acknowledgements:} 

\noindent
We gratefully acknowledge partial 
support through the TMR-network "Finite Temperature Phase Transitions in
Particle Physics", EU contract no. ERBFMRX-CT97-0122.

\end{document}